\documentclass{article}

\usepackage{PRIMEarxiv}

\usepackage[utf8]{inputenc} 
\usepackage[T1]{fontenc}    
\usepackage{hyperref}       
\usepackage{url}            
\usepackage{booktabs}       
\usepackage{amsfonts}       
\usepackage{nicefrac}       
\usepackage{microtype}      
\usepackage{lipsum}
\usepackage{fancyhdr}       
\usepackage{graphicx}       
\usepackage{amsmath}
\usepackage{amsfonts}
\graphicspath{{media/}}     

\title{Electromagnetic Effective-Degree-of-Freedom Limit of a MIMO System in {2-D} Inhomogeneous Environment}

\author{Shuai S. A. Yuan $^{1}$, Zi He $^{2}$, Sheng Sun $^{3}$, Xiaoming Chen $^{4}$, Chongwen Huang $^{1}$ and Wei E. I. Sha $^{1,}$*\\ \\
$^{1}$ College of Information Science and Electronic Engineering, Zhejiang University, Hangzhou 310027, China.\\
$^{2}$ School of Electrical Engineering and Optical Technique, Nanjing University of Science and Technology, \\Nanjing 210094, China\\
$^{3}$  School of Electronic Science and Engineering, University of Electronic Science and Technology of China, \\Chengdu 611731, China.\\
$^{4}$ School of Information and Communications Engineering, Xi'an Jiaotong University, Xi'an 710049, China.\\ \\
\texttt{Correspondence: weisha@zju.edu.cn}}

\begin{document}
\maketitle
\begin{abstract}
Compared with a single-input-single-output (SISO) wireless communication system, the benefit of multiple-input-multiple-output (MIMO) technology originates from its extra degree of freedom (DOF), also referred as scattering channels or spatial electromagnetic (EM) modes, brought by spatial multiplexing. When the physical sizes of transmitting and receiving arrays are fixed, and there are sufficient antennas (typically with half-wavelength spacings), the DOF limit is only dependent on the propagating environment. Analytical methods can be used to estimate this limit in free space, and some approximate models are adopted in stochastic environments, such as Clarke's model and Ray-tracing methods. However, this DOF limit in an certain inhomogeneous environment has not been well discussed with rigorous full-wave numerical methods. {In this work, volume integral equation (VIE) is implemented for investigating the limit of MIMO effective degree of freedom (EDOF) in three representative two-dimensional (2-D) inhomogeneous environments. Moreover, we clarify the relation between the performance of a MIMO system and the scattering characteristics of its propagating environment.}
\end{abstract}

\keywords{MIMO system \and Degree of freedom \and Inhomogeneous Green's function \and Volume integral equation \and Correlation}

\section{Introduction}
Based on the Shannon's information theory \cite{shannon}, multiple-input-multiple-output (MIMO) technology in wireless communications has achieved great success over the recent twenty years \cite{telatar1999capacity}, along with the emerging massive \cite{TL2014}, reconfigurable-intelligent-surface (RIS) \cite{cui2020, Huang2019} MIMO and holographic \cite{Pizzo2020, Huang2020} MIMO systems. As electromagnetic (EM) wave is the physical carrier of information, the capacity bound of a space-constrained MIMO system from EM perspective deserves to be investigated for exploring and examining the emerging MIMO technologies. 

The degree of freedom (DOF) of a MIMO system refers to the rank (number of significant eigenvalues) of its correlation matrix \cite{tse2005fundamentals}, or the number of scattering channels \cite{Bucci} or spatial EM modes \cite{miller2019waves}, which shows a clear physical meaning. DOF indicates the benefit brought by MIMO technology and is frequently used for characterizing the performance of a MIMO system. A more convenient figure of merit used in the MIMO system is the effective degree of freedom (EDOF) \cite{muharemovic2008antenna, Migliore2006, Shuai2021}, which approximately characterizes its equivalent number of single-input-single-output (SISO) systems. The EDOF is directly related to the capacity and easy for calculation. 

The DOF/EDOF limit has been discussed in several research works, which can be roughly classified into mode-based and channel-based analyses. For the mode-based analysis, the DOF limit is intuitively calculated by counting the available EM modes in a constrained spatial or angular domain, including the DOF limit between two planes in free space \cite{miller2000communicating}, between one plane and half-space\cite{ShuaiPra} and in Rayleigh environment \cite{loyka2004information, Pizzo2020,Davide2020}. {For the channel-based analysis, the channel matrices are constructed with some empirical or EM methods, then the DOF/EDOF can be obtained through singular value decomposition (SVD), including the MIMO channels based on traditional models \cite{kronecker,clarke1968statistical} (Clarke, Kronecker, etc.), EM Green's function \cite{piestun2000electromagnetic, MD2019, Wallace2008, wen2007multi}, full-wave simulation \cite{JR2006, Mats2021}, and dyadic Green's function in free space \cite{Shuai2021}.} Particularly, the communication channels have been investigated with various EM methods, such as the numerical parabolic equations for tunnels \cite{Sarris2016}, stochastic Green's function for stochastic environments \cite{Peng2020}, ray-tracing methods \cite{RT2016} and some approximate models for the environment with finite scatterers \cite{Burr2003,Bentosela2014,Poon2006,Hanpinitsak2017}. {These works provide some useful results for the DOF/EDOF of MIMO systems in free space and stochastic environments. Nevertheless, for some representative inhomogeneous environments, the impacts of their scattering characteristics on the performance of a MIMO system are not well investigated with numerical methods. }

In this work, mainly three contributions are made. {First, we provide an explicit deduction and explanation of the EM foundation of a MIMO system. Second, three representative numerical examples are given by using the two-dimensional (2-D) volume integral equation (VIE), the MIMO performances in these environments have not been discussed with EM numerical methods before. Third, the relation between the performance of a MIMO system and the scattering characteristics of the 2-D inhomogeneous environment is clarified.} This paper demonstrates a simple but representative application of computational electromagnetic methods in wireless communications. The proposed model is particularly useful for evaluating and exploring the performance limit of MIMO systems in various deterministic environments which cannot be well modeled by empirical methods.
\section{Methodology}
\subsection{EM model for analyzing EDOF limit}
{If there are amounts of scatterers in the propagating environment, the strengths of signals going through this environment will follow a statistical model due to the central limit theorem, and the effect of specific physical properties, such as the size and shape of scatterers, are insignificant. These statistical models are generally applicable to various scenarios, which largely simplifies the modeling task of a communication system. However, for a few of scatterers, these models are not applicable, and the scattering characteristics of scatterers need to be taken into account with rigorous numerical methods.}

\begin{figure}[ht!]
	\centering
	\includegraphics[width=3.3in]{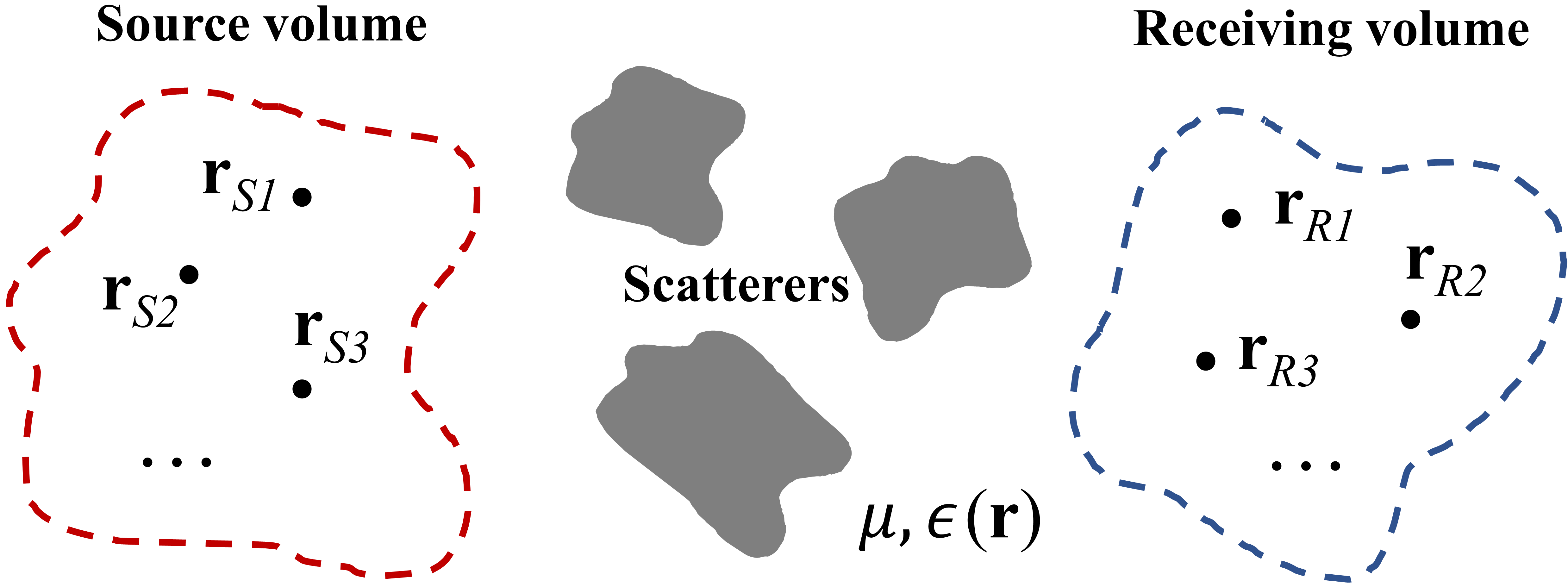}
	\caption{A general model of a MIMO system in isotropic inhomogeneous environment. Transmitting antennas at the positions $\mathbf{r}_{S n}\left(n=1, \ldots, N_{S}\right)$ are distributed in the source volume, and receiving antennas at the positions $\mathbf{r}_{R m}\left(m=1, \ldots, N_{R}\right)$ are distributed in the receiving volume. The permeability of the environment is $\mu$ and the permittivity is $\epsilon(\mathbf{r})$ characterizing arbitrary scatterers.}
	\label{MIMO system}
\end{figure}
We consider a general MIMO model in isotropic inhomogeneous environment, with homogeneous permeability $\mu$ and inhomogeneous permittivity $\epsilon(\mathbf{r})$ characterizing arbitrary scatterers, as depicted in Fig. 1. A set of $N_S$ transmitting antennas at the positions $\mathbf{r}_{S n}\left(n=1, \ldots, N_{S}\right)$ are distributed in the source volume, $N_R$ receiving antennas at the positions $\mathbf{r}_{R m}\left(m=1, \ldots, N_{R}\right)$ are distributed in the receiving volume. The transmitting and receiving antennas are modeled as isotropic point sources/receivers, which is a widely-used approach for the DOF/EDOF analysis \cite{miller2019waves, Shuai2021}.

With source function $\mathbf{J}(\mathbf{r^{\prime}})$, the resulting electric field is
\begin{equation}
\mathbf{E}\left(\mathbf{r}\right)=-j \omega \mu \int \bar{\mathbf{G}}_{in}\left(\mathbf{r}, \mathbf{r}^{\prime}\right) \mathbf{J}(\mathbf{r^{\prime}})d \mathbf{r^{\prime}},
\end{equation}
where $\bar{\mathbf{G}}_{in}$ is the inhomogeneous dyadic Green's function satisfying \cite{chew1995waves}
\begin{equation}
\nabla \times  \nabla \times \bar{\mathbf{G}}_{in}\left(\mathbf{r}, \mathbf{r}^{\prime}\right)-\omega^{2}\mu \epsilon(\mathbf{r}) \bar{\mathbf{G}}_{in}\left(\mathbf{r}, \mathbf{r}^{\prime}\right)= \bar{\mathbf{I}} \delta\left(\mathbf{r}-\mathbf{r}^{\prime}\right),
\end{equation}
with $\bar{\mathbf{I}}$ the unit tensor. The matrix representation of $\bar{\mathbf{G}}_{in}$ is
\begin{equation}
\bar{{\mathbf{G}}}_{in}=\left[\begin{array}{ccc}
G_{xx} & G_{xy}  & G_{xz} \\
G_{yx} & G_{yy}  & G_{yz} \\
G_{zx} & G_{zy}  & G_{zz}
\end{array}\right],
\end{equation}
\noindent where each element is a scalar inhomogeneous Green's function corresponding to one polarization of field and one polarization of source, denoted by its subscript. {Under point source/receiver approximation, a channel matrix can be built for each scalar Green's function.} For $G_{xy}$, a set of $N_S$ point sources at the positions $\mathbf{r}_{S n}$ in the source volume are excited with the complex amplitudes $t_{yn}$ along the $\emph{y}$ polarization, then the superposed $\emph{x}$-polarized electric field $f_{xm}$ generated at the positions $\mathbf{r}_{Rm}$ in the receiving volume would be
\begin{equation}
f_{xm}=\sum_{n=1}^{N_{S}} G_{xy}\left(\mathbf{r}_{R m}, \mathbf{r}_{Sn}\right) t_{yn}=\sum_{n=1}^{N_{S}} h_{mn} t_{yn},
\end{equation}
where $h_{mn}=G_{xy}\left(\mathbf{r}_{R m}, \mathbf{r}_{Sn}\right)$ is the scalar inhomogeneous Green's function with the pre-designed source and receiving positions. If $ t_{yn}$ and $f_{xm}$ are collected in the two column vectors ${t_y}=[t_{y1}, t_{y2}, \dots, t_{yN_S}]^T$ and ${f_x}=[f_{x1}, f_{x2}, \dots, f_{xN_R}]^T$, we can define the projection from the point sources to the point receivers as
\begin{equation}
{f_x}=\mathbf{H}_{xy} {t_y},
\end{equation}
with the channel matrix
\begin{equation}
\mathbf{H}_{xy}=\left[\begin{array}{cccc}
h_{11} & h_{12} & \cdots & h_{1 N_{S}} \\
h_{21} & h_{22} & \cdots & h_{2 N_{S}} \\
\vdots & \vdots & \ddots & \vdots \\
h_{N_{R} 1} & h_{N_{R} 2} & \cdots & h_{N_{R} N_{S}}
\end{array}\right].
\end{equation}
Similarly, for full polarizations in Eq. (4), the complex amplitudes of sources ${t}=[{t}_x {t}_y {t}_z]^T=[t_{x1}, \cdots, t_{xN_S}, t_{y1}, \cdots, t_{yN_S}, t_{z1}, \cdots, t_{zN_S}]^T$ is a 3$N_S$ $\times$ 1 vector, and the complex amplitudes of received signals ${f}=[{f}_x {f}_y {f}_z]^T=[f_{x1}, \cdots, f_{xN_R}, f_{y1}, \cdots, f_{yN_R}f_{z1}, \cdots, f_{zN_R}]^T$ is a 3$N_R$ $\times$ 1 vector. The two column vectors are then related by a 3$N_R$ $\times$ 3$N_S$ EM channel matrix
\begin{equation}
\mathbf{H}=\left[\begin{array}{ccc}
\mathbf{H}_{xx} & \mathbf{H}_{xy}  & \mathbf{H}_{xz} \\
\mathbf{H}_{yx} & \mathbf{H}_{yy}  & \mathbf{H}_{yz} \\
\mathbf{H}_{zx} & \mathbf{H}_{zy}  & \mathbf{H}_{zz}
\end{array}\right],
\end{equation}
\par\noindent where the nine $N_R$ $\times$ $N_S$ matrices correspond to the nine scalar inhomogeneous Green's functions. The three polarizations of electric field are orthogonal, hence, we could write the channel of full polarizations in a matrix form without loss of information. Notice that the (one-sample) correlation matrix $ \mathcal{R} ={\mathbf{H}}{\mathbf{H}}^\dagger$ or $ {{\mathbf{H}}^\dagger\mathbf{H}}$ ($^\dagger$ is the Hermitian operator) sharing the same eigenvalues is used for estimating the performance of a MIMO system, as the transmitting (receiving) powers are related by {$\bar{\mathbf{G}}_{in}^\dagger\bar{\mathbf{G}}_{in}$ ($\bar{\mathbf{G}}_{in}\bar{\mathbf{G}}_{in}^\dagger$).} 

The performance of a MIMO system can be characterized by the communication capacity, DOF and EDOF. Although the capacity is an ultimate demand, we are focusing on the eigenvalues of the correlation matrix in this work, i.e., influence of propagating environment, and the signal-to-noise ratio (SNR) can be considered as a constant. EDOF is used here for convenient calculation and comparison, as observing the significant eigenvalues for DOF is tedious. For a MIMO system, EDOF represents its equivalent number of SISO systems, and can be calculated by \cite{muharemovic2008antenna}
\begin{equation}\label{DoF}
\Psi_e\left(\mathcal{R}\right)=\left(\frac{\operatorname{tr}\left(\mathcal{R}\right)}{\left\|\mathcal{R}\right\|_{F}}\right)^{2}=\frac{\left(\sum_{i} \sigma_{i}\right)^{2}}{\sum_{i} \sigma_{i}^{2}},
\end{equation}
\begin{figure}[ht!]
	\centering
	\includegraphics[width=2.5in]{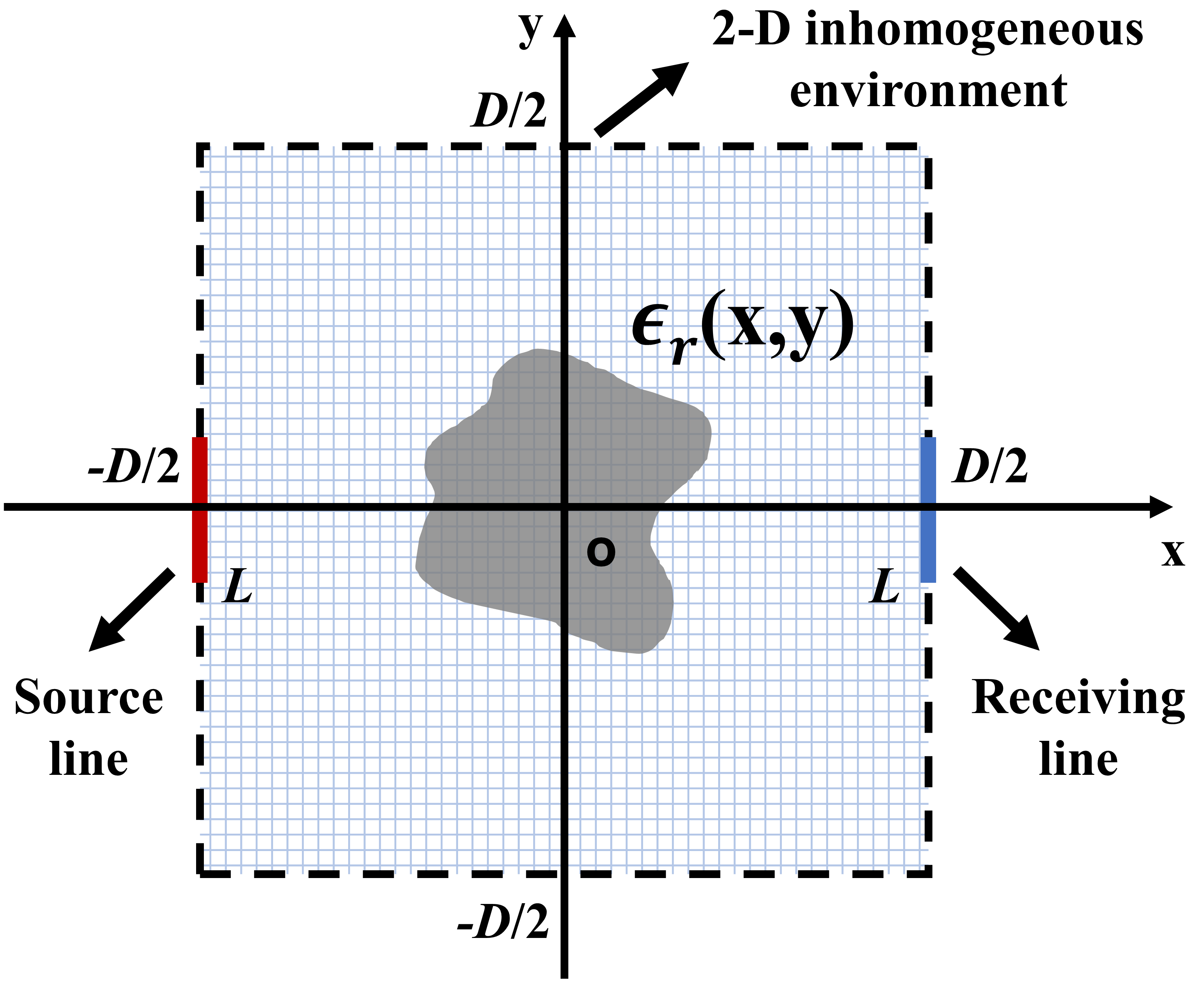}
	\caption{A 2-D inhomogeneous environment for the EDOF analyses. The red and blue lines are the source and receiving lines with the same length $L$, and $2L/\lambda_0+1$ point sources/receivers (slightly smaller than 0.5$\lambda_0$ spacings, $\lambda_0$ is the free-space wavelength) are uniformly distributed along the source/receiving lines for approaching the EDOF limit, and $D$ is the distance between the source and receiving lines. The blue grid represents the discretized grids, and $\epsilon_{r}(x,y)$ characterizes arbitrary isotropic scatterers. }
	\label{MIMO system}
\end{figure}
\noindent where $\operatorname{tr}(\cdot)$ represents the trace operator, the subscript $F$ denotes the Frobenius norm and $\sigma_{i}$ are the eigenvalues of $\mathcal{R}$. This equation has been proved to be sufficiently accurate for estimating the performance of a MIMO system\cite{Shuai2021}. Generally, the EDOF limit is determined by both the size of source/receiving volume and the propagating environment \cite{piestun2000electromagnetic, MD2019, Wallace2008}, and the latter is of our concern here. With sufficient antennas (typically with half-wavelength spacings), the EDOF limit of a space-constrained MIMO system would be 1 in free space at far field (line of sight), and reach the maximum value in Rayleigh environment (rich scattering). When some scatterers are placed into the propagating environment to form an inhomogeneous environment, we will observe an increase of EDOF, which can be used for characterizing different inhomogeneous environments. 
\subsection{Numerical method for inhomogeneous Green's function}
Based on Eq. (5), when there are sufficient antennas, the performance limit of a space-constrained MIMO system is completely determined by the inhomogeneous Green's function of the propagating environment, which can be investigated with full-wave numerical methods. As a proof of concept, also for a low computational cost, the 2-D VIE for the TM wave (only $\emph{z}$ polarization in Fig. 2) is used for getting some fundamental results and gaining useful physical conclusions. Besides, various full-wave methods can be readily applied to corresponding scenarios following the framework proposed in Eqs. (4-8). {For the TM wave case, the dielectric material is represented by equivalent polarization current ${{J}_z}(x, y)$ in VIE \cite{peterson1998computational}}
\begin{equation}
{{J}_z}(x, y)=j \omega \epsilon_{0}\left[\epsilon_{r}(x, y)-1\right] {E}_{z}(x, y),
\end{equation}
where $E_z(x, y)$ is the total electric field, $\epsilon_{0}$ is the permittivity in free space, and $\epsilon_{r}$ is the relative permittivity. Then, the scattering problem can be modeled by
\begin{equation}
E_{z}^{\mathrm{inc}}(x, y)=\frac{J_{z}(x, y)}{j \omega \varepsilon_{0}\left(\varepsilon_{r}(x, y)-1\right)}+j \omega \mu_{0} A_{z}(x, y),
\end{equation}
where
\begin{equation}
A_{z}(x, y)=\iint J_{z}\left(x^{\prime}, y^{\prime}\right) \frac{1}{4 j} H_{0}^{(2)}(k_0 R) d x^{\prime} d y^{\prime},
\end{equation}
and
\begin{equation}
R=\sqrt{\left(x-x^{\prime}\right)^{2}+\left(y-y^{\prime}\right)^{2}},
\end{equation}
$k_0$ is the free-space wavenumber, and $H_{0}^{(2)}$ is the 2-D Green's function in free space. {After implementing spatial discretization in Fig. 2 ($N$ cells), we can use the 2-D pulse basis}
\begin{equation}
p_n(x, y)= \begin{cases}1 & \text { if }(x, y) \in \text { cell } n \\ 0 & \text { otherwise }\end{cases},
\end{equation}
{the polarization current can be expressed as}
\begin{equation}
J_z(x, y) \cong \sum_{n=1}^N j_n p_n(x, y).
\end{equation}
{Substituting the discrete current into the Eq. (10) yields}
\begin{equation}
E_z^{\text {inc }}(x, y) \cong \sum_{n=1}^N j_n\left(\frac{\eta p}{j k\left[\varepsilon_r\right.}\left(\frac{(x, y)}{x, y)-1]}+j k \eta \iint_{\text {cell } n} \frac{1}{4 j} H_0^{(2)}(k R) d x^{\prime} d y^{\prime}\right)\right.,
\end{equation}
{which can be written into a matrix form}
\begin{equation}
\mathbf{E}_{z}^{\mathrm{inc}}=\bar{\mathbf{Z}}\mathbf{J_z},
\end{equation}
{with the entries}
\begin{equation}
\left[\begin{array}{c}
E_2^{\text {inc }}\left(x_1, y_1\right) \\
E_2^{\text {inc }}\left(x_2, y_2\right) \\
\cdot \\
\cdot \\
\cdot \\
E_z^{\text {inc }}\left(x_N, y_N\right)
\end{array}\right]=\left[\begin{array}{cccc}
Z_{11} & Z_{12} & \cdots & Z_{1 N} \\
Z_{21} & Z_{22} & & Z_{2 N} \\
\cdot & \cdot & & \\
\cdot & \cdot & & \\
\cdot & \cdot & & \\
Z_{N 1} & Z_{N 2} & \cdots & Z_{N N}
\end{array}\right]\left[\begin{array}{c}
j_1 \\
j_2 \\
\cdot \\
\cdot \\
\cdot \\
j_N
\end{array}\right],
\end{equation}
{where}
\begin{equation}
Z_{m n}=\frac{k \eta}{4} \iint_{\mathrm{cell} n} H_0^{(2)}\left(k R_m\right) d x^{\prime} d y^{\prime} \quad m \neq n,
\end{equation}
\begin{equation}
Z_{m m}=\frac{\eta}{j k\left(\varepsilon_{r m}-1\right)}+\frac{k \eta}{4} \iint_{\text {cell } m} H_0^{(2)}\left(k R_m\right) d x^{\prime} d y^{\prime},
\end{equation}
\begin{equation}
R_m=\sqrt{\left(x_m-x^{\prime}\right)^2+\left(y_m-y^{\prime}\right)^2}.
\end{equation}
The inverse of matrix $\bar{\mathbf{Z}}$ is accelerated by the conjugate-gradient fast-fourier-transform (CG-FFT) method \cite{Sarkar, zhang2001three}. By using the VIE, we can quickly calculate the field distribution excited by a point source at any position in arbitrary environments, thus obtain the columns of the channel matrix. In fact, 2-D TM cases would be sufficient for gaining some fundamental EM insights, as many useful channel models are 2-D models, e.g., Clarke's model \cite{clarke1968statistical}. 

For the specific setups, as shown in Fig. 2, the distance between the source and receiving lines is $D$, the lengths of them are both $L$, and $2L/\lambda_0+1$ source/receiving points are uniformly distributed along the source/receivng lines (slightly smaller than 0.5$\lambda_0$ spacings), which is sufficient for approaching the EDOF limit of a space-constrained MIMO system \cite{Pizzo2020}. Arbitrary scatterers can be denoted by $\epsilon_{r}(x,y)$, for metallic structures, we use the complex permittivity of copper at 10 GHz ($\epsilon_{r}=1-1.044\times 10^8$j). The above would be the parameter settings for the following numerical examples.

\section{Numerical examples}
\subsection{Key-hole scenario}
\begin{figure}[ht!]
	\centering
	\includegraphics[width=3.4in]{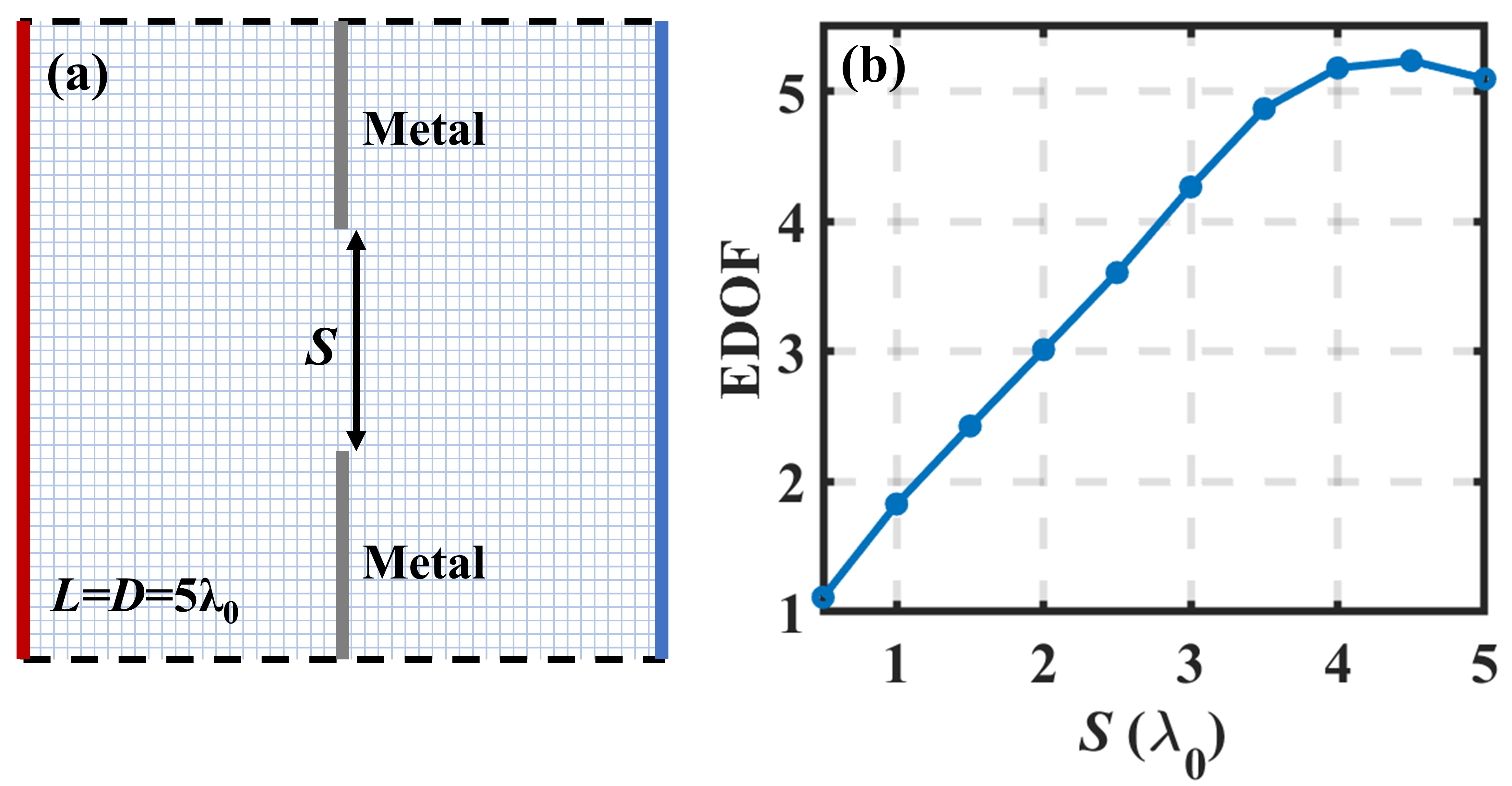}
	\caption{EDOF limit in key-hole scenario. (a) Diagram of the key-hole scenario. $L=5\lambda_0$ is the length of source/receiving line, and $D=5\lambda_0$ is the distance between the source and receiving lines. A metallic sheet, with a $S \lambda_0$ hole in the center, is placed in the middle between the source and receiving lines. (b) EDOF limit versus $S$.}
	\label{MIMO system}
\end{figure}
\begin{figure}[ht!]
	\centering
	\includegraphics[width=3.4in]{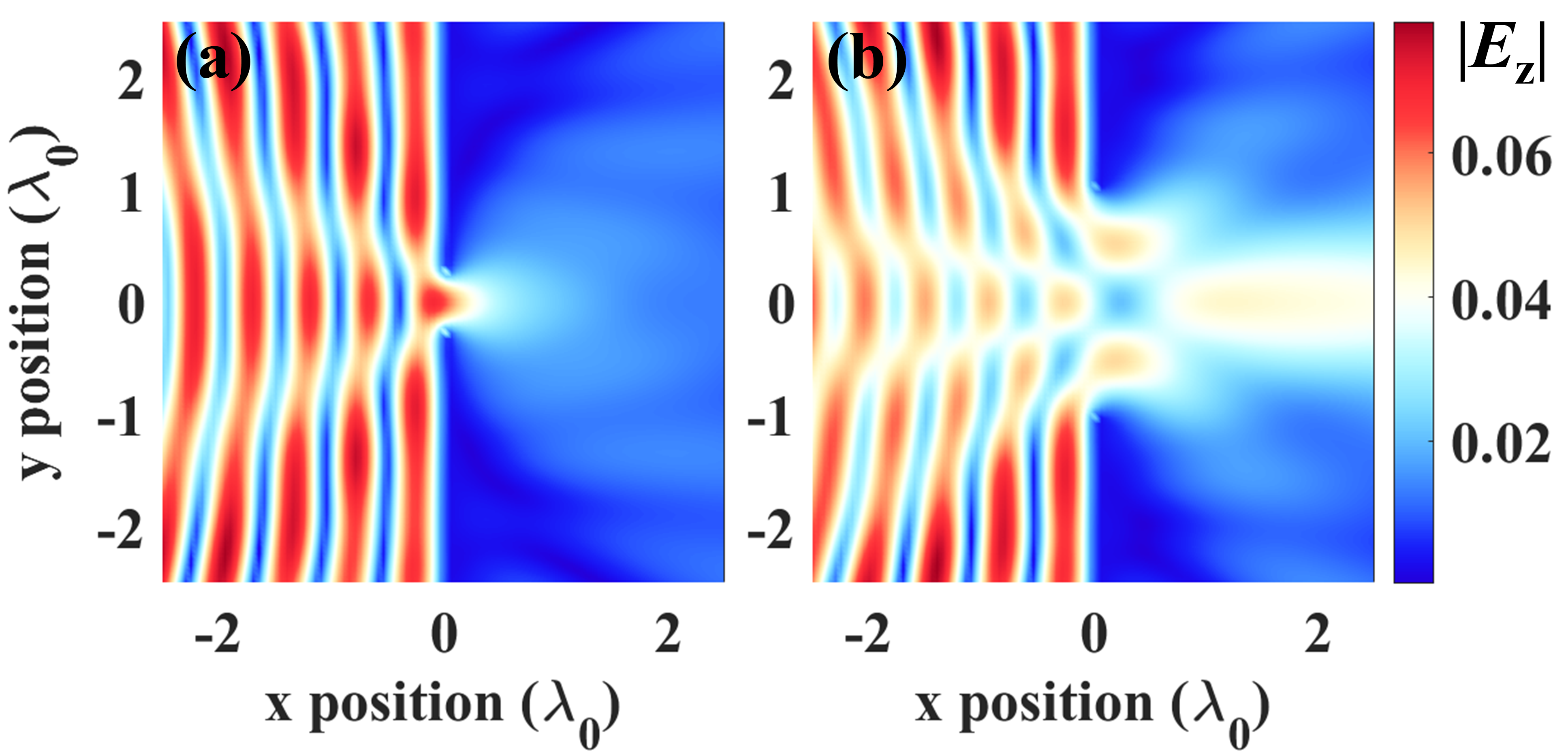}
	\caption{Electric field $|E_z|$ excited by a point source at position (-2.5,0). (a) $S=0.5\lambda_0$. (b) $S=2\lambda_0$.}
	\label{MIMO}
\end{figure}
Key-hole scenario is a special and well-known MIMO system, where the rank of the correlation matrix is very low due to the particular propagating environment. Some approximate methods and experiments have been conducted for investigating this scenario \cite{chizhik2002keyholes, almers2006keyhole}. Here we present some results with the proposed model. The key-hole scenario discussed is shown in Fig. 3(a). A large metallic sheet with an aperture is set in the middle between the source and receiving lines, which becomes a key-hole scenario when the size of aperture ($S$) is very small. The lengths of the source and receiving lines are set relatively long ($L=D=5\lambda_0$), so that there will still be EDOF gain when the sheet is removed. The relation between $S$ and EDOF is depicted in Fig. 3(b). It can be observed that the EDOF will be gradually reduced to 1, i.e., become equivalent to that of a SISO system, when the size of hole keeps decreasing. Also, the electric fields $|E_z|$ excited by a point source at position (-2.5,0) with different $S$ are demonstrated in Fig. 4. When $S=0.5\lambda_0$, few power could pass through the hole, and the field at the receiving line is rather uniform, which is not good for the MIMO communication. More signals can be received and the field is more focusing at the receiving line when $S=2\lambda_0$.
\subsection{Cylindrical scatterers}
\begin{figure}[ht!]
	\centering
	\includegraphics[width=3.4in]{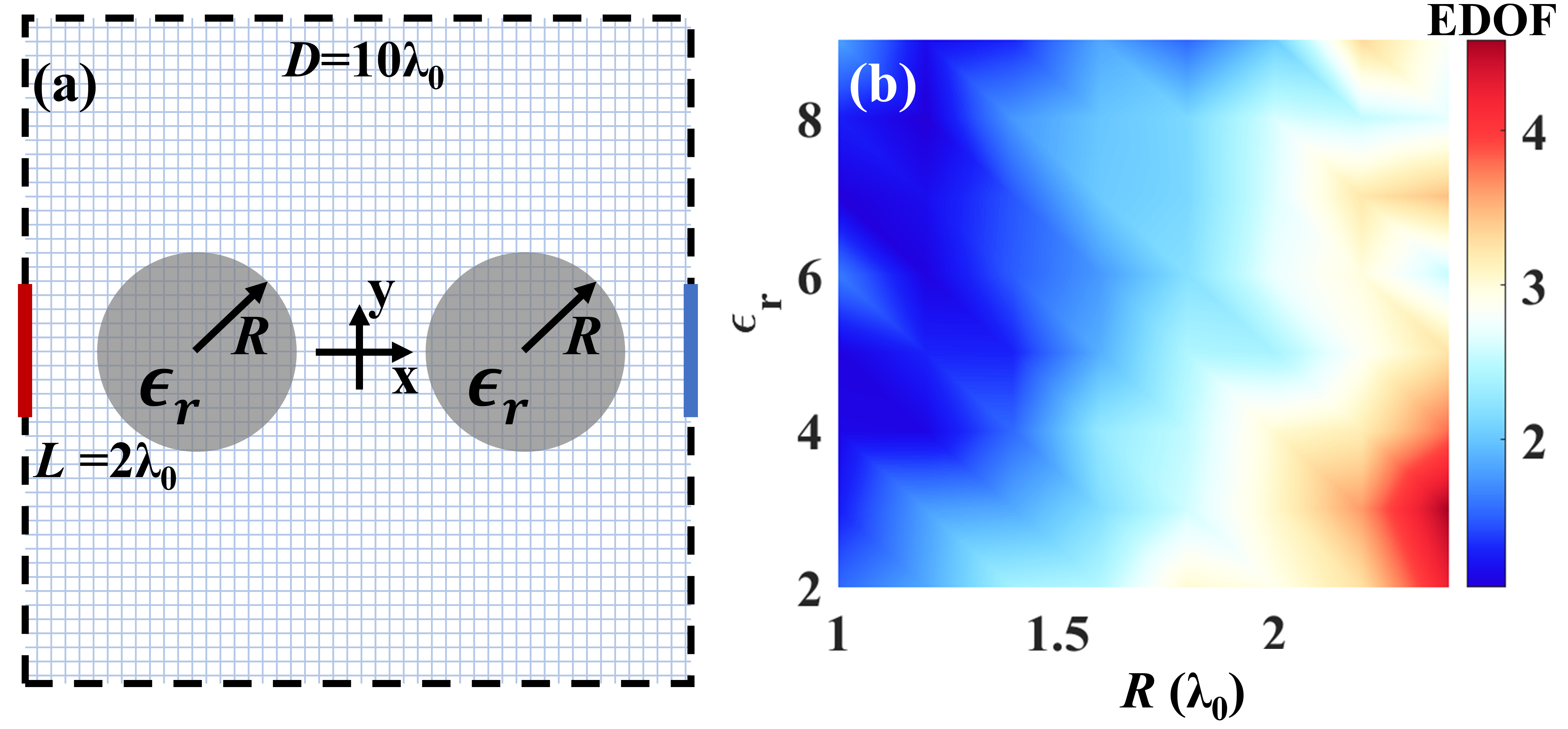}
	\caption{EDOF limit in inhomogeneous environment with double cylindrical scatterers. (a) Diagram of the double-scatterer environment. The lengths of source and receiving lines are both $L=2\lambda_0$, and the distance between them is $D=10\lambda_0$. Two identical dielectric cylinders, with the radius $R$ and relative permittivity $\epsilon_{r}$, are placed in the left and right quadrisection points between the source and receiving lines. (b) EDOF limit versus $R$ and $\epsilon_{r}$.}
	\label{MIMO}
\end{figure}
\begin{figure}[ht!]
	\centering
	\includegraphics[width=3.4in]{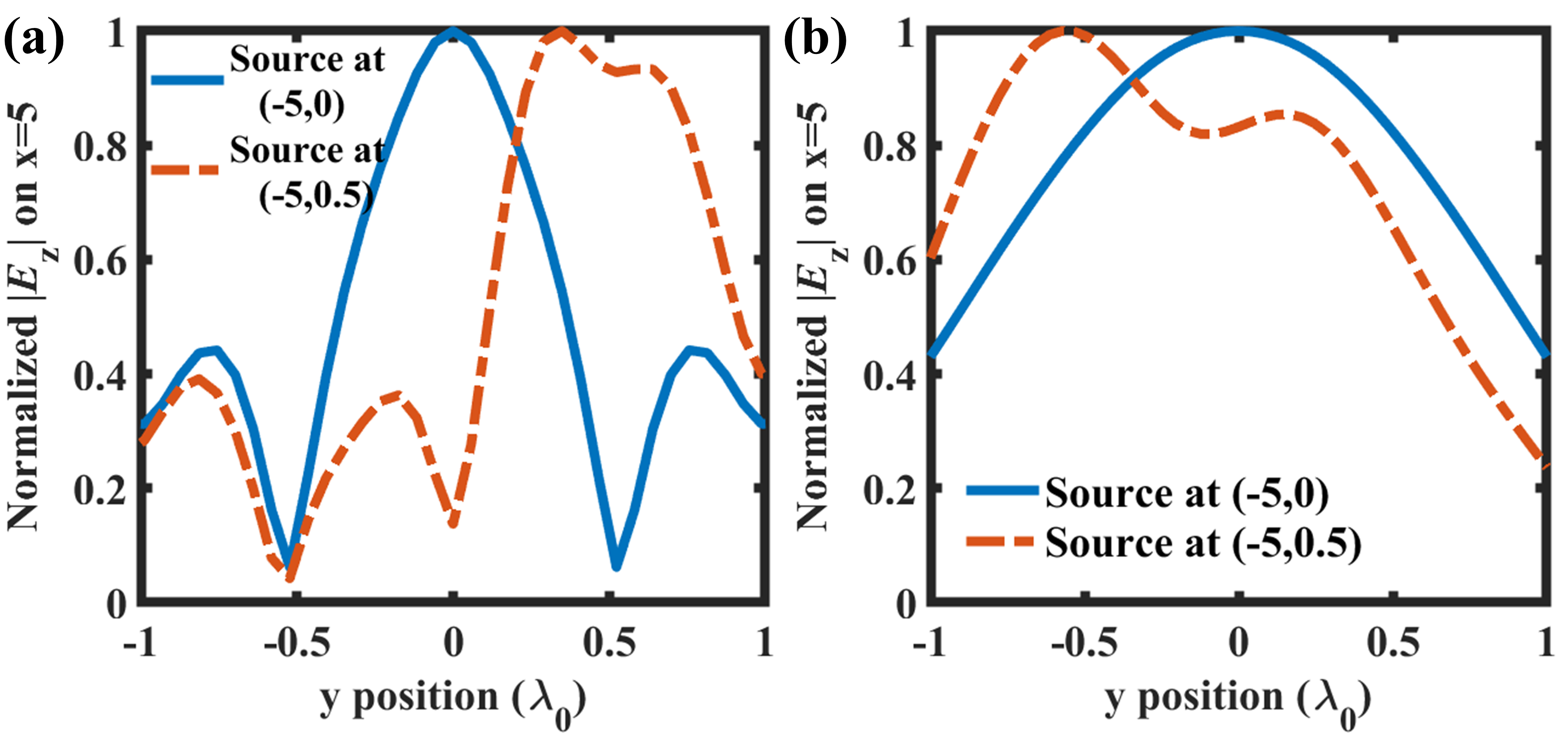}
	\caption{{Normalized $|E_z|$ at the receiving line (at $x=5$) excited by the two point sources at the positions (-5,0) and (-5,0.5). (a) $\epsilon_{r}=3$, $R=2.4\lambda_0$. (b) $\epsilon_{r}=6$, $R=2\lambda_0$.}}
	\label{MIMO}
\end{figure}
The EDOF limit in inhomogeneous environments constructed by 2-D cylindrical scatterers is also investigated. Here, we demonstrate a typical case after testing several kinds of combinations, including single, double, and uniform or random array of cylindrical scatterers. It has been found that the two-scatterer case depicted in Fig. 5(a) can achieve relatively higher EDOF, where the two dielectric cylinders, with the same radius $R$ and relative permittivity $\epsilon_{r}$, are placed in the left and right quadrisection points between the source and receiving lines. The lengths of source and receiving lines are $L=2\lambda_0$ and the distance between them is $D=10\lambda_0$, which fulfills the far-field condition (EDOF = 1 without the scatterers). 

The EDOF versus $R$ and $\epsilon_{r}$ is shown in Fig. 5(b), from which we can observe that larger $R$ can significantly improve EDOF, while the influence of $\epsilon_{r}$ is not distinct. At the maximum value of EDOF (EDOF $=$ 4.3), the cylindrical scatterers provide fairly good MIMO performance even close to that in a canonical Rayleigh channel (EDOF $=$ 4.5). {The normalized electric fields $|E_z|$ excited by point sources at positions (-5,0) and (-5,0.5) in the best case ($R=2.4\lambda_0$, $\epsilon_{r}=3$) are depicted in Fig. 6(a), and that in the worse case ($R=2\lambda_0$, $\epsilon_{r}=6$) are depicted in Fig. 6(b). Apparently, the fields in the former case are more focusing, leading to a lower field correlation (inner product between the received electric fields excited by the two point sources at two different positions), thus a higher EDOF.}
\subsection{Cavity structure}
\begin{figure}[ht!]
	\centering
	\includegraphics[width=3.4in]{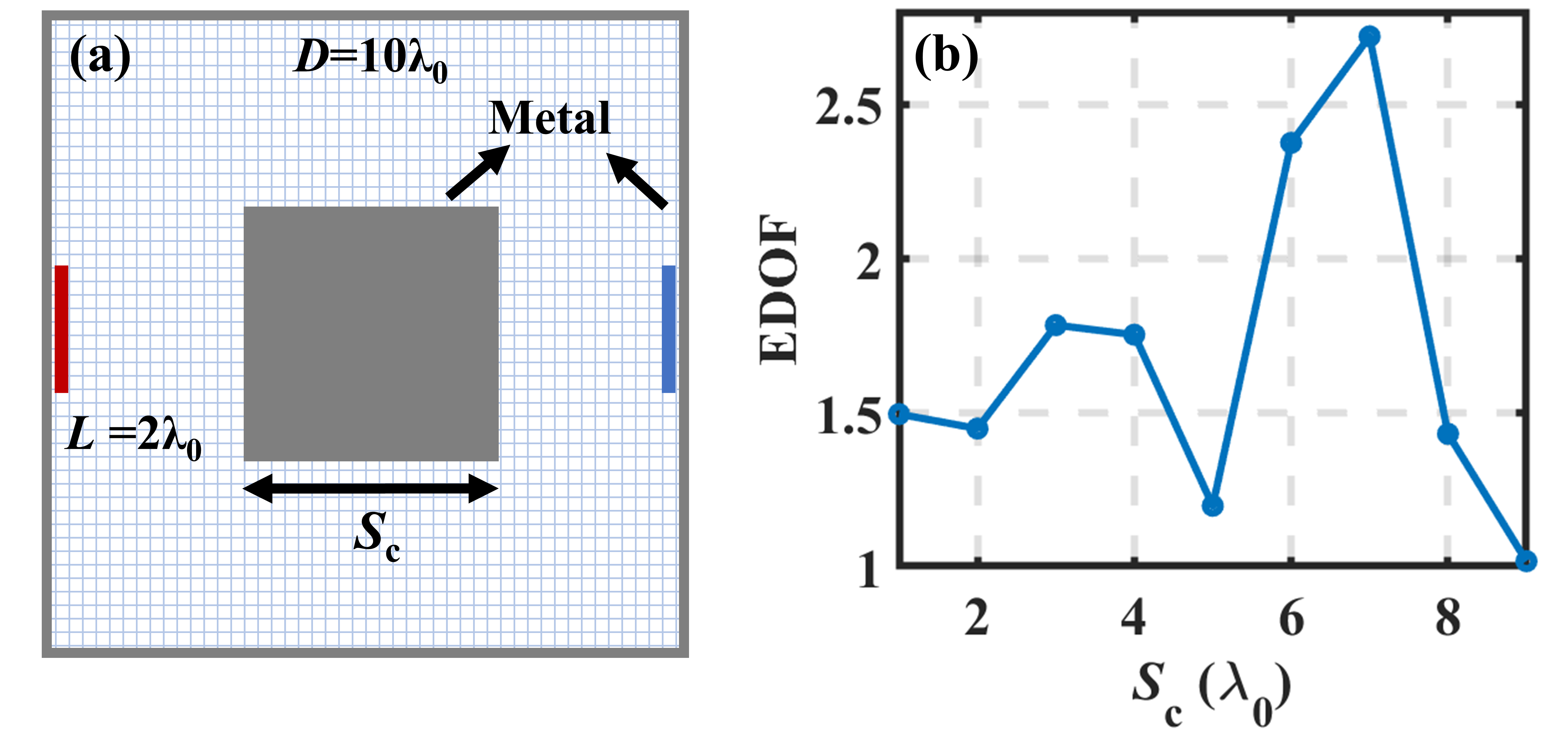}
	\caption{EDOF limit in a metallic cavity structure. (a) Diagram of the cavity. The lengths of source and receiving lines are both $L=2\lambda_0$, and the distance between them is $D=10\lambda_0$, and the side length of cavity is slightly larger than $D$. A metallic square obstacle, with the side length $S_c$, is placed in the center between the source and receiving lines. (b) EDOF limit versus $S_c$.}
	\label{MIMO}
\end{figure}
\begin{figure}[ht!]
	\centering
	\includegraphics[width=3.4in]{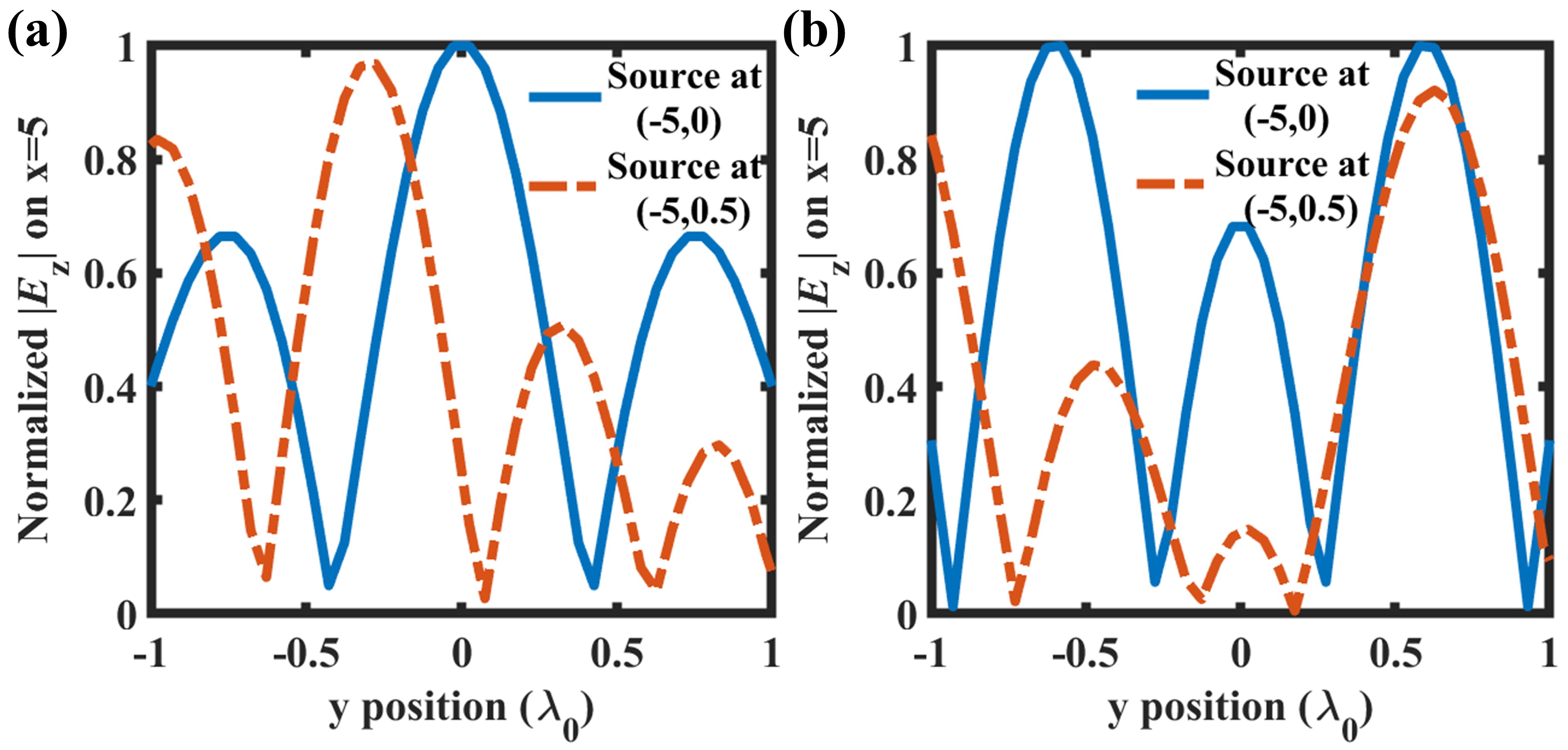}
	\caption{{Normalized $|E_z|$ at the receiving line (at $x=5$) excited by the two point sources at the positions (-5,0) and (-5,0.5). (a) $S_c=7\lambda_0$. (b) $S_c=5\lambda_0$.}}
	\label{MIMO}
\end{figure}
As depicted in Fig. 7(a), in a metallic cavity, a metallic square obstacle with a side length $S_c$ is placed in the middle between the source and receiving lines to enhance the multi-path effect \cite{Sarkar2019}. The lengths of source and receiving lines are $L=2\lambda_0$ and the distance between them is $D=10\lambda_0$ (the same as the parameters in the cylindrical case). The EDOF versus $S_c$ is shown in Fig. 7(b), it can be observed that a relatively larger obstacle can enhance multi-path effect, while the EDOF will approach 1 if the propagating path is largely blocked by an oversized obstacle, just similar to the key-hole case. {The decrease at $S_c=5\lambda_0$ is on account of the oscillation of spatial correlation, e.g., Clarke's model.} {The normalized electric fields $|E_z|$ excited by the two point sources at the positions (-5,0) and (-5,0.5) in the best case ($S_c=7\lambda_0$) are depicted in Fig. 8(a), and that in the worst case ($S_c=5\lambda_0$) are depicted in Fig. 8(b). The field correlations in the two cases are 12 and 32, respectively. Different from the cylindrical case, the correlations are hard to be directly compared from the figures, which means that the MIMO performance in a complicated environment still needs to be estimated with a strict model rather than an intuitive method.}
\section{Relation between MIMO performance and scattering characteristics}
The MIMO performance is determined by its correlation matrix $\mathcal{R}$, where each entry $\mathcal{R}_{mn}$ is the inner product between the received fields generated by the $m$th source and the $n$th source. A good MIMO performance requires that $\mathcal{R}_{mn}$ is kept as small as possible when $m\ne n$. This requirement indicates that two types of fields could produce ideal MIMO performances: the Dirac-function type of field and the Gaussian-white-noise type of field, as $\mathcal{R}_{mn}$ is 0 when $m\ne n$ in the two situations. In the followings, we are going to discuss what kinds of inhomogeneous environments could produce the fields close to the two ideal situations.

For the Dirac-function type of field, using a 2-D cylindrical scatterer is naturally a good route, as shown in Fig. 9. A large cylinder has the scattering characteristic of strong forward scattering, thus producing a focusing field. Also, because of its angular symmetry, the fields excited by the $m$th source and the $n$th source have almost the same patterns but a spatial translation, as illustrated in the right part of Fig. 9. Therefore, the inner product between the fields ($\mathcal{R}_{mn}$) will be smaller if the fields are more focusing, as the case in Fig. 6. In fact, this scenario is quite similar to the Clarke's model for the Rayleigh channel. Rayleigh channel can be simulated by the superposition of uniformly-arrived plane waves \cite{clarke1968statistical}, and its electric field distribution in Cartesian coordinate is
\begin{figure}[ht!]
	\centering
	\includegraphics[width=2.35in]{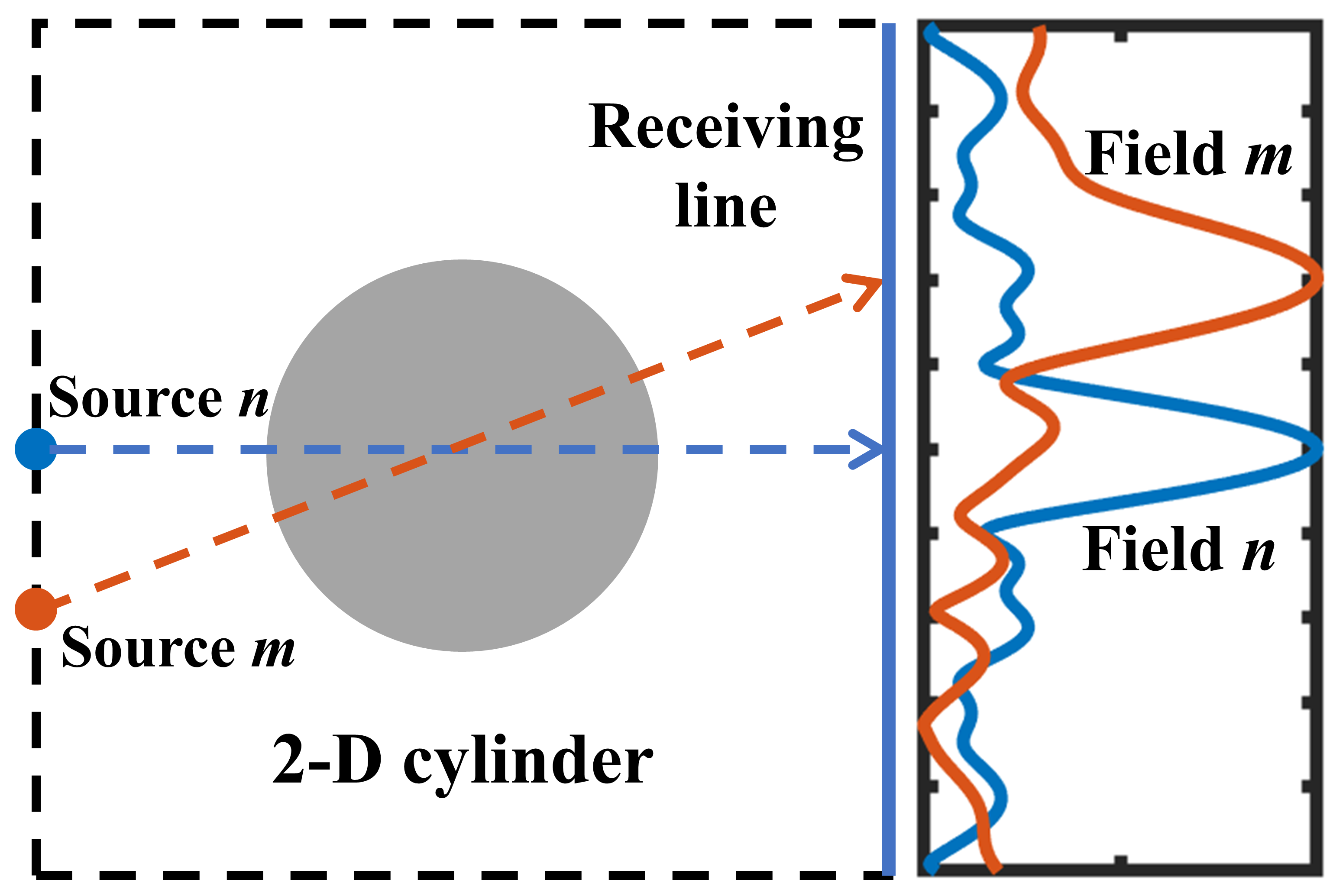}
	\caption{Diagram of the electric fields excited by two point sources at different positions, with a 2-D cylindrical scatterer in the center.}
	\label{MIMO}
\end{figure}
\begin{equation}
E_{z}(x,y)=\frac{1}{\sqrt{N}} \sum_{n=1}^{N}\exp\left\{-j \varphi_{n}\right\},
\end{equation}
where $N$ is the number of plane waves, $\varphi_{n}=k_0\cos\alpha_nx+k_0\sin\alpha_ny$, $\alpha_n$=$\pi n/N$ denote the directions of plane waves, and $1/\sqrt{N}$ is the average of power. When $N$ waves arrive from any directions with equal probability, the autocorrelation function of the electric field is
\begin{equation}
\begin{aligned}
R_{E_{z}}(\xi)&= \left\langle E_{z}(x,y)E_{z}^{*}(x+\xi,y)\right\rangle _{\rm{av}}\\
&=\frac{1}{{N}}\sum_{n=1}^{N} \left\langle \exp \left\{jk_0  \cos \alpha_{n}\xi\right\}\right\rangle _{\rm{av}}\\
&=J_0(k_0\xi),
\end{aligned}
\end{equation}
where $^{*}$ denotes the conjugate, $\left\langle\cdot\right\rangle_{\rm{av}}$ represents the average, $\xi$ is the distance between the two receivers and $J_0$ is the zero-order Bessel function. Obviously, the correlation in the Rayleigh channel is just the average of phase delays between the two receivers, where the incident plane waves arrive from $\alpha_{n}$ with the same probability. The deterministic fields $n$ and $m$ in Fig. 9 can be regarded as the effective samples (realizations) for random $E_{z}(x,y)$ and $E_{z}(x+\xi,y)$ in Eq. (16), respectively, indicating that a large 2-D cylindrical scatterer in the TM case could be a physical analogy of the Rayleigh channel.

Hence, for a deterministic propagating environment, large and angular-symmetric scatterers will produce good MIMO performances when the receivers are properly positioned, which is also tenable for the three-dimensional full-polarization cases. For highly-irregular scatterers, the influences of phases and polarizations are complex, and thus it is hard to draw some general conclusions. Rigorous numerical methods are needed for modeling and understanding these complicated environments, and the challenge of computational cost can be eased by making some reasonable approximations, such as parabolic equations for tunnels and ray-tracing methods for stations \cite{Sarris2016}. Moreover, for Eq. (3), we also tried Born series \cite{born2013principles} for getting some insights of multiple-scattering effects. However, the Born series is not suitable for communication problems because the series is not convergent for the large-scale high-contrast problems.

For the Gaussian-white-noise type of field, the correlations need to be considered from the perspectives of statistical properties. For example, in a reverberant chamber for creating various stochastic environments, the received fields would appear noise-like properties due to the stirring and time average \cite{Vahldieck2005, Chen2012}. Large amounts of scatterers in the propagating environment always produce this kind of field. However, these are stochastic channels and out of our main concerns in this work.

\section{Conclusion}
{In this work, VIE is combined with an EM MIMO model for estimating the EDOF limit in 2-D inhomogeneous environments. Three representative numerical examples are presented, which provides insights for exploring the EDOF limit of deterministic MIMO systems. The theoretical framework could readily be implemented in various communication scenarios with corresponding numerical algorithms incorporating full polarizations. The proposed results are useful for understanding the EM information theory and the designs of MIMO antenna array.}

\vspace{6pt}


%


	
	
	\bibliographystyle{unsrt}  
	\bibliography{Bibliography}
\end{document}